# Decreased Thermal Conductivity of Polyethylene Chain Influenced by Short Chain Branching

Danchen Luo,[1] Congliang Huang,[1,2,a)] Zun Huang[1]

[1] School of Electrical and Power Engineering, China University of Mining and Technology, Xuzhou 221116, China.

[2] Department of Mechanical Engineering, University of Colorado, Boulder, Colorado 80309-0427, USA.

a) Author to whom correspondence should be addressed. E-mail: huangcl@cumt.edu.cn.

**Abstract**

In this paper, we have studied the effect of short branches on the thermal conductivity of a polyethylene (PE) chain. With a reverse non-equilibrium molecular dynamics method applied, thermal conductivities of the pristine PE chain and the PE-ethyl chain are simulated and compared. It shows that the branch has a positive effect to decrease the thermal conductivity of a PE chain. The thermal conductivity of the PE-ethyl chain decreases with the number density increase of the ethyl branches, until the density becomes larger than about 8 ethyl per 200 segments, where the thermal conductivity saturates to be only about 40% that of a pristine PE chain. Because of different weights, different types of branching chains will cause a different decrease of thermal conductivities, and a heavy branch will leads to a lower thermal conductivity than a light one. This study is expected to provide some fundamental guidance to obtain a polymer with a quite low thermal conductivity.

**Keyword**: thermal conductivity, polymer, branching chain, molecular dynamics simulation.

# 1. Introduction

Not only a high thermal conductivity (TC) but also a quite low TC are desired for polymers because of their wide applications[1-6], such as high TC for application as the thermal interface material[7,8] and low TC for application as thermal insulation material. Single polymer chains and highly aligned polymer fibers have attracted a wide attention due to their potential high TC. [9-16] Although a single polymer chain may possess a high TC, polymers are generally regarded as thermal insulators because of their very low thermal conductivities on the order of 0.1 W·m$^{-1}$·K$^{-1}$ [17]. One of the reasons for the low TC is that the polymer chains are randomly coiled in the polymers, which effectively shortens the mean free path (MFP) of heat-carrying phonons.[18,19] Another reason is that the TC of these polymers can be significantly influenced by the morphology of individual chains.[14-17,20-22] Besides these two reasons, the method to further decrease the TC of a polymer is still desired to develop a thermal insulators.

There have already been some methods to reduce the TC of a polymer chain. Liao et al. [23] tuned the TC of a polymer chain by atomic mass modifications and found that heavy substituents hinder heat transport substantially. Robbins and Minnich [16] found that even perfectly crystalline polynorbornene has an exceptionally low thermal conductivity near the amorphous limit due to extremely strong anharmonic scattering. Most recently, Ma and Tian [24] studied the influence of the side chains on the thermal conductivity of bottlebrush polymers, and predicted that side chains dominate the heat conduction and could lead to a lower TC. Some other studies also shown that chain segment disorder, or the random rotations of segments in a chain, will lead to a lower TC. [15,25-28]

In this paper, we take the effect of branches into account to probe a way to reduce the TC of a polymer. Considering the complex structure of a polymer, we just focus on the polyethylene (PE) chain with branches. Results turn out that the TC of a PE chain with branches can be decreased to be only 40% that of a pristine chain. It can be predicted that, if the chain in the polymer is branched with short chains, the TC of a PE polymer can be decreased a lot. The paper is organized as follows: firstly, a reverse non-equilibrium molecular dynamics (RNEMD) method is introduced; and then the effects of backbone chain length, branching chain location, branching chain type, and the number density of branching chains are investigated. This study is expected to provide some fundamental guidance to obtain a polymer with a quite low TC.

## 2. Simulation Method

The software package BIOVIA Materials Studio is applied to build the initial configuration of the single extended PE chain and the modified PE chain, and then to simulate the TC. A single PE chain is established by replicating the PE chain segments which is the unit cell of PE's idealized bulk lattice structure with a cross-sectional area of 18 Å$^2$, a length of 2.507Å. A pristine PE chain and a PE chain with a branching ethyl (PE-ethyl for short) are shown in Fig. 1. After the structure of the PE chain built, we firstly optimize the structure before carrying out molecular dynamic simulations. Considering that the condensed-phase optimized molecular potentials for atomistic simulation studies (COMPASS II) [29-31] was created to accurately simulate the structural, vibrational, and thermo-physical properties of PE in isolated and condensed phases[32,33], and it has already been successfully applied to study thermal transport, [11,15,34] the COMPASS II potential is also applied in this paper. Before calculating the TC, we firstly relax the system in an NVT (constant number of atoms, temperature, and volume) ensemble at a temperature of 300 K for 125 ps. And then, a NVE (constant number of atoms, volume, and energy) ensemble is applied to release the thermal stress. The Nose-Hoover thermostat [35,36] is applied for obtaining a constant temperature. We double-check that the total energy has reached minimum and become unchangeable at the end of NVT (or NVE) ensemble to make sure that our systems have already been equilibrated.

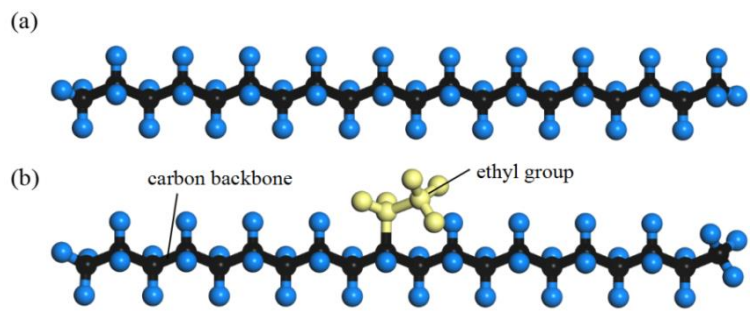

*Fig. 1 Structure of PE chains used in the simulation: (a) A single extended PE chain with a chain length of about 10 segments; (b) a PE chain with a branching ethyl.*

For calculating TC, the RNEMD [37,38] simulation is performed on the well equilibrated structures. In the RNEMD method, each of the simulation boxes is divided into several slabs with a periodic boundary applied in the heat transfer direction. As that shown in Fig. 2, the simulation system is divided into several slabs (20 to 200 slabs, depending on the chain length), slab 0 is the "hot" slab, and the slab N/2 is the "cold" slab. Other slabs are used to obtain the temperature distributions. The heat flux is created by exchanging velocities of particles in "cold" and "hot" slabs. The cold slab donates its "hottest" particles (particles with the highest kinetic energy) to the hot slab in exchange for the latter's "coolest" particles (particles with the lowest kinetic energy). Performing this exchange periodically results in the heating up of the hot slab and cooling down of the cold slab. This process eventually yields a steady-state temperature gradient due to thermal conduction through slabs separating the cold and hot slabs. The thermal conductivity is calculated exactly by the relationship,

$$\lambda = -\frac{\sum \frac{m}{2}(v_h^2 - v_c^2)}{2tA\langle \partial T/\partial Z \rangle} \quad (1)$$

where the sum is taken over all transfer events during the simulation time $t$, $m$ is the mass of the atoms, $v_c$ and $v_h$ are the velocities of the identical mass particles that participate in the exchange procedure from the cold and hot slabs, respectively. $A$ is the average cross-sectional area which is calculated by surface area (also called accessible surface area calculated by the Connolly surface model) [39,40] divided by chain length, here the chain length is the number of segments multiplied by the length of the unit segment 2.507Å. Cross-sectional areas of the pristine PE chain, PE-ethyl chain, and others are listed in the Table 1, which possess a mean value of 14.705 Å$^2$ with a branch-caused uncertainty less than 1.6%. Such a small difference in cross-sectional area will not lead to a large thermal conductivity difference as that caused by branches (discussed later). The thermal conductivity present in our work could be scaled by a different choice of cross-sectional area for comparison. With a time step of 1 fs, a total simulation time 0.1 ns is taken to get a good linear temperature distribution. With heat flux printed out every 0.1 ps, the TC is calculated at the last step. The temperature distribution of a simulation with a length of 100 segments is shown in Fig. 2 as an example. The linear temperature region is fitted to obtain the temperature gradient for the calculation of the effective TC by using the Fourier's law. The TC calculated at different simulation times is shown in *Fig. 3*. It shows that 0.1ns is long enough to get a converged TC.

Table 1 Cross-sectional areas of the pristine PE chain, PE-ethyl chain, etc.

|  | 50 segments | 75 segments | 100 segments |
| --- | --- | --- | --- |
| Pristine PE | 14.502 Å$^2$ | 14.469 Å$^2$ | 14.539 Å$^2$ |
| PE-ethyl | 14.872 Å$^2$ | 14.717 Å$^2$ | 14.697 Å$^2$ |
| PE-beneze | 14.718 Å$^2$ | 14.902 Å$^2$ | 14.890 Å$^2$ |
| PE-phenoxy | 14.858 Å$^2$ | 14.750 Å$^2$ | 14.822 Å$^2$ |
| PE-ethoxy | 14.717 Å$^2$ | 14.848 Å$^2$ | 14.730 Å$^2$ |
| PE-methoxy | 14.741 Å$^2$ | 14.729 Å$^2$ | 14.601 Å$^2$ |
| PE-ethylene | 14.769 Å$^2$ | 14.650 Å$^2$ | 14.671 Å$^2$ |
| PE-hydroxy | 14.524 Å$^2$ | 14.649 Å$^2$ | 14.561 Å$^2$ |

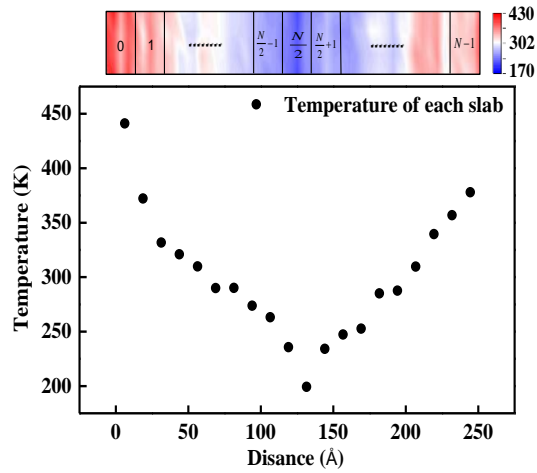

*Fig. 2 Temperature distribution of a single extended PE chain with a length of 100 segments (25 nm).*

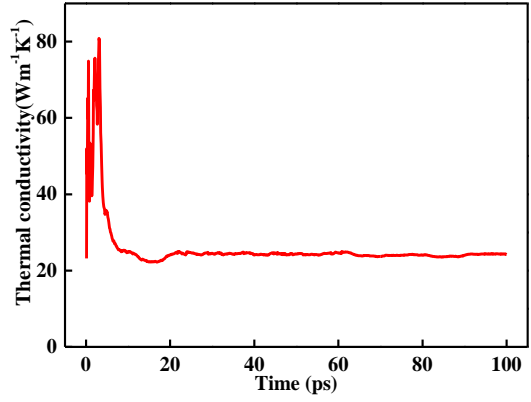

*Fig. 3 TC calculated at different simulation times of a pristine PE chain with a length of 100 segments.*

A quantum correction is sometimes applied to rectify the result predicted by a molecular dynamic method which includes no quantum effects. Although applying the quantum corrections to the classical molecular dynamic predictions does not bring them into a better agreement with the quantum predictions compared to the uncorrected classical molecular dynamic, [41] a quantum correction is still given here for future comparisons. In molecular dynamic simulations, the temperature $T_{MD}$ is calculated based on the mean kinetic energy of the system. By assuming that the total system energy is twice the mean kinetic energy at $T_{MD}$ and equals to the total phonon energy of the system at the quantum temperature $T_q$, with the Debye density of states [42], correction is made through, [43,44]

$$T_{\mathrm{MD}} = \frac{3\hbar}{\kappa_B} \int_0^{\omega_D} \frac{\omega^3}{\omega_D^3} \left[ \frac{1}{e^{\hbar\omega/k_B T_q}-1} + \frac{1}{2} \right] d\omega \tag{2}$$

where $\hbar$ is reduced Planck's constant, $\kappa_B$ is Boltzmann's constant, and $\omega$ is the phonon frequency, $\omega_D$ is the Debey frequency. With the quantum-corrected temperature 85 K which corresponds to $T_{MD}$ = 300 K [9] substituted in Eq. (2), $\omega_D$ could be calculated from Eq. (2). With the $\omega_D$ value substituted in Eq. (2), the relation between $T_{MD}$ and $T_q$ is obtained and shown in Fig. 4(a). Then, the quantum rectified thermal conductivity can be calculated by, [45]

$$k_q = \left(\frac{dT_{MD}}{dT_q}\right) k \tag{3}$$

where $k$ is the thermal conductivity presented in this paper, $dT_{MD}/dT_q$ is calculated from Eq. (2) and values are shown in Fig. 4 (b).

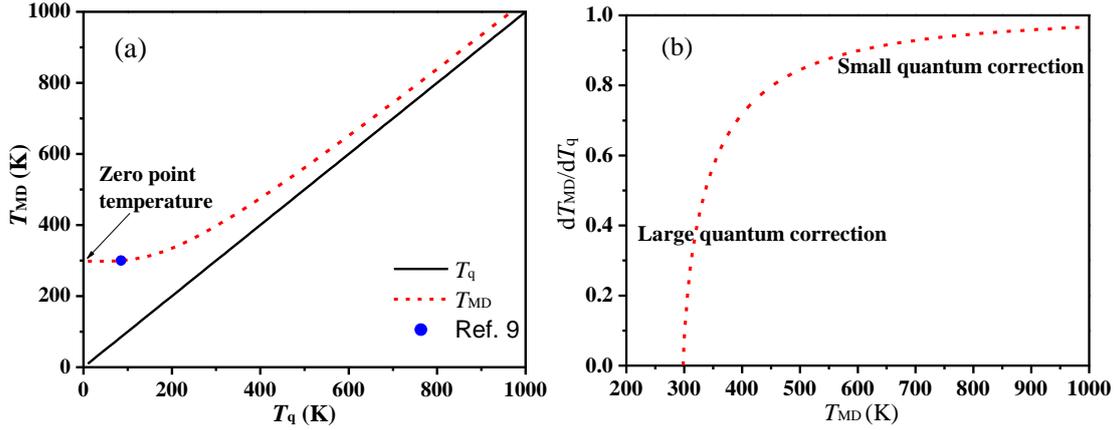

*Fig. 4 Quantum rectification: (a) MD temperature versus quantum temperature; (b) ratio of MD to quantum temperature versus MD temperature.*

## 3. Results and discussions

Firstly, the length dependence of the TC of a pristine PE chain is investigated and compared with that in early researches. And then, the TC of the pristine PE chains with different length is compared with that of a PE-ethyl chain. Thirdly, the effect of the branch arrangements is considered. Finally, the influence of the branching chain types and the number density of branching chains are taken into account.

### 3. 1 Length dependence of TC

TCs of the pristine PE chain with different chain lengths at 300 K are firstly simulated, and presented in Fig. 5(a). Results in some previous research [14,46,47] about the pristine PE chains are also added in Fig. 5(a) for comparison. As that shown in Fig. 5(a), there is an obvious increase of the TC with the increase of the chain length. Even with the length increasing to be 200 nm, the TC still not converges, which suggests that some portion of the phonons can still travel ballistically in such a length. Our simulation work confirms that the TC of a pristine PE chain will increase with the increasing number of segments (or chain length), and the TC of a single PE chain is several orders of magnitude larger than that of a PE polymer. In Fig. 5(a), the TC difference between different works should be attributed to the different simulation methods, considering that the NEMD method is applied in our and Hu et al.'s work and EMD is applied in the work of Ni et al. and Liu and Yang. It seems that a NEMD method will give a higher TC than an EMD method. This was also noticed in other studies[48,49] and some explanations can be found there.

According to the Boltzmann transport equation and the Matthiessen's rule, there is a linear relationship between the inverse of the TC ($1/k$) and the inverse of the sample length ($1/L$).[15] After plotting $1/k$ against $1/L$ in Fig. 5(b), we obtain the intrinsic TC of an infinitely long pristine chain by linearly extrapolating the data to $1/L = 0$ when $L = \infty$ [see Fig. 5(b)], and the TC comes out to be 303 W·m$^{-1}$·K$^{-1}$. This value also agrees well with that in Ref. [9]. All these confirm the reliability of the RNEMD method.

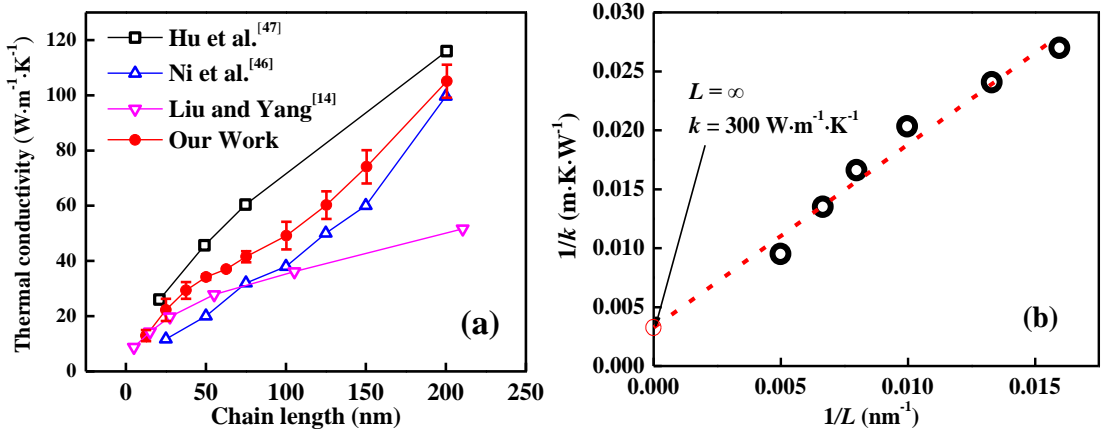

*Fig. 5 TC of a pristine PE chain: (a) compared with results simulated by Ni et al. [46], Hu et al.[47] and Liu and Yang[14]. (b) Inverse of the TC of the single PE chain plotted against the inverse of the chain length, showing a linear dependence. Extrapolation of the dashed line leads to the TC of an infinitely long chain.*

TCs of the pristine PE chain and the PE-ethyl chain with lengths ranging from 100 to 500 segments (or 25.07-125.35 nm) are compared in Fig. 6. It turns out that both TCs of the pristine PE chain and the PE-ethyl chain increase with the increasing length, and the TC of a PE-ethyl chain is only about 75% that of a pristine PE chain. For illustrating the underlying mechanism of the lower TC of the PE-ethyl chain, the vibrational density of states (VDOS) is calculated by using the Fourier transform of the velocity autocorrelation function. Results are compared between the pristine PE chain and the PE-ethyl chain with 50 segments, as shown in Fig. 7. Considering the low-frequency (< 20THz) phonons dominate the TC due to their high group velocities and long mean free paths, [23] the lower VDOS of the PE-ethyl chain in the low frequency should be responsible for the lower TC, where the branch acts as a center of low-frequency-phonon scattering.

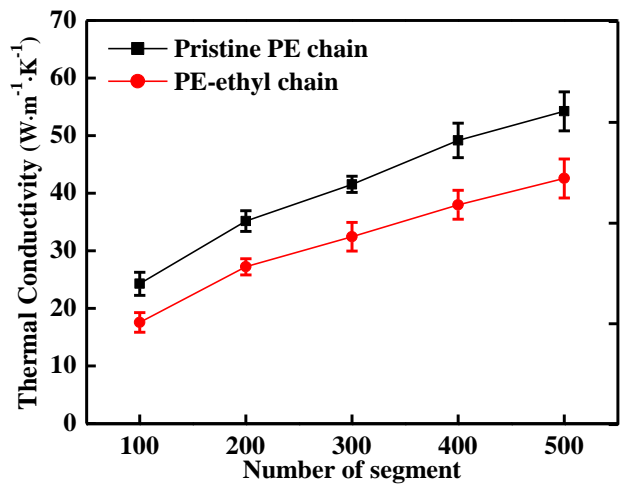

*Fig. 6 Length dependence of the TC of the PE chain and the PE-ethyl chain.*

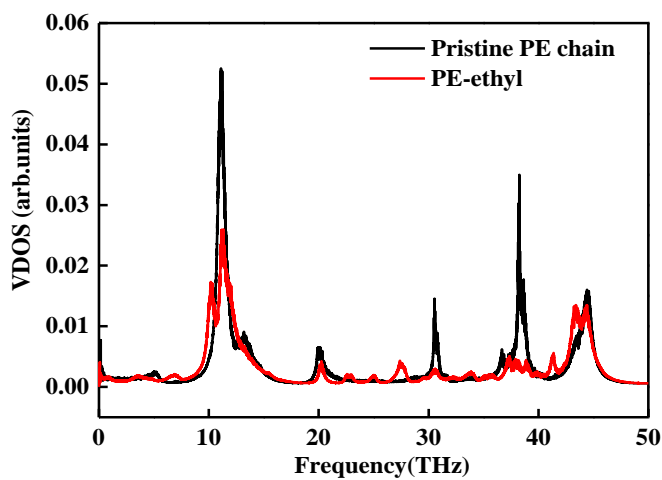

*Fig. 7 VDOS of PE chains*

### 3.2 Influence of branch arrangements

The influence of branch locations is considered in this part. For a pristine PE chain, there are different locations from the simulation region boundary to the branching ethyl. Five special locations are selected to branch a short chain, labeled as P1, P2, P3, P4 and P5 respectively, as that shown in Fig. 8(a). The result in Fig. 8(b) confirms that the presence of a branching chain can truly reduce the TC, and the average TC of a PE-ethyl chain is about 0.7 times that of a pristine PE chain. Our simulations also indicate that there is almost a similar thermal conductivity for different branch locations in Fig. 8(b). This is attributed to the periodic boundary conditions applied in the simulation. The small discrepancies of the TC between different locations should be caused by the different distance of the branch from the simulation boundary. If the boundary and the branching chains are both thought as defects on a pristine PE chain, the TC with ethyl located at the middle of the chain (P1) will be lower than other TCs (P2, P3, P4 and P5), because of the small distance from the middle of the chain to the system boundary. This is confirmed by results in Fig. 8(b).

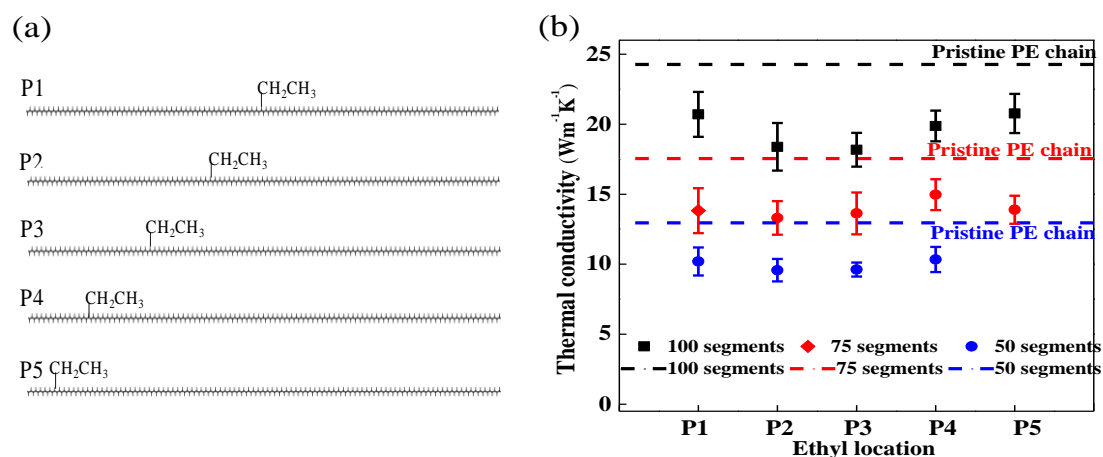

*Fig. 8 TC of the PE-ethyl chain with different branch locations: (a) 100-segment structures used in the simulation; (b) Effect of branch locations on the TC. Dashed lines stand for the pristine PE chains.*

**3. 3 Influence of branching chain types and number density of branching chains**

Seven different types of short chains are branched on the middle segment of a PE chain for comparisons. They are different from the weight and the type of chemical bonds between backbone and the branching chains, as shown in Fig. 9, which are listed as phenoxy group, phenyl group, ethoxy group, methoxy group, ethyl group, ethylene group, and hydroxy group respectively. The black column, red column and blue column in Fig. 9 stand for different chain length. The relative masses of different branches are also shown in Fig. 9. We can see that all types of branching chains lead to a decrease of TC, and a heavy branch leads to a lower TC than a light one, except for the ethylene group in which TC may be further decreased by a different bond. It agrees with the conclusions in Ref. [23] that a chain modified by a heave atom possesses a lower thermal conductivity than that modified by a light one, where the modifying atom can be thought a special short branch. We conclude that different kinds of branching chains will lead to a different decrease of thermal conductivities because of the different weight. More studies are still needed to probe the effect of bonds between the backbone and the branch on TC.

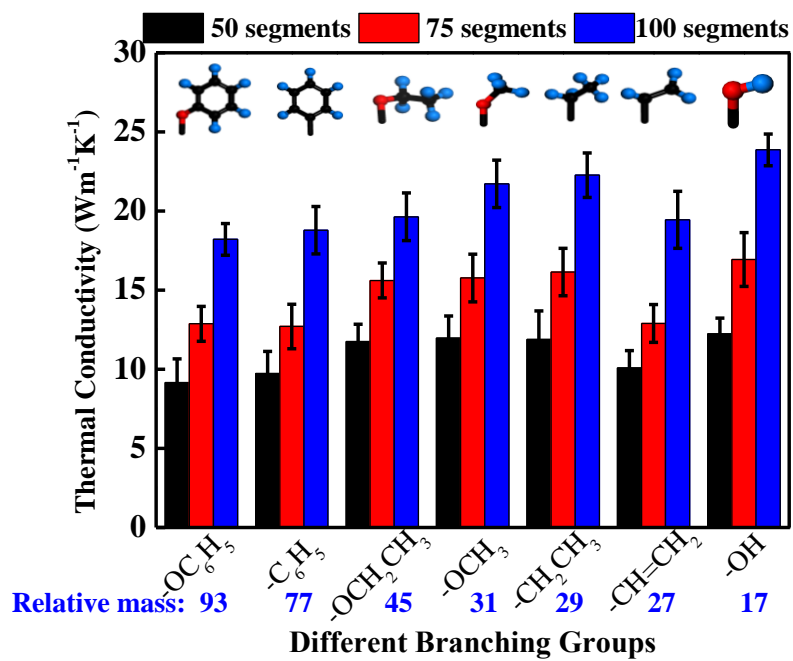

Fig. 9 TCs of PE chains with different types of branching chains.

The effect of the number density of branches is studied in this part. The number density of branches is defined as the number of branches divided by the number of PE segments. 200 segments (50.14 nm) are applied as a periotic unit in the simulation, and the ethyl group is selected as the branch. Considering there are different locations on a PE chain to branch an ethyl group, we only consider two special location arrangements, i. e., the aligned arrangement and the fork arrangement, as that shown in Fig. 10(a). For the aligned arrangement of 10 branching ethyl, they are equally distributed on the PE chain, only a part of the chain is shown in Fig. 10(a); for the fork arrangement of 10 branching ethyl, every 2 branching ethyl are located at the same segment of the PE chain, as that shown in Fig. 10(a). The corresponding TC of these two arrangements is shown in Fig. 10(b). It shows that a larger number density of branches leads to a lower TC for both arrangements. With an increase of the number density of branches, the TC of a PE-ethyl chain converges to be only 40% that of the pristine PE chain. This can be understood by that with the increase of the number density, the distance between branches is reduced, and the long-MFP phonons will be decreased until the TC converges to a constant value. It can be predicted that if the PE-ethyl chain instead of the pristine PE chain is used to build up a polymer, the TC of the polymer will be much reduced, because of the lower TC of the PE-ethyl chain and additional masses of branches. A polymer composed of what kind of chains will possess a lower TC is the key point of this paper, and more studies are still needed to figure out the effect of long branches on the TC of a pristine chain [24].

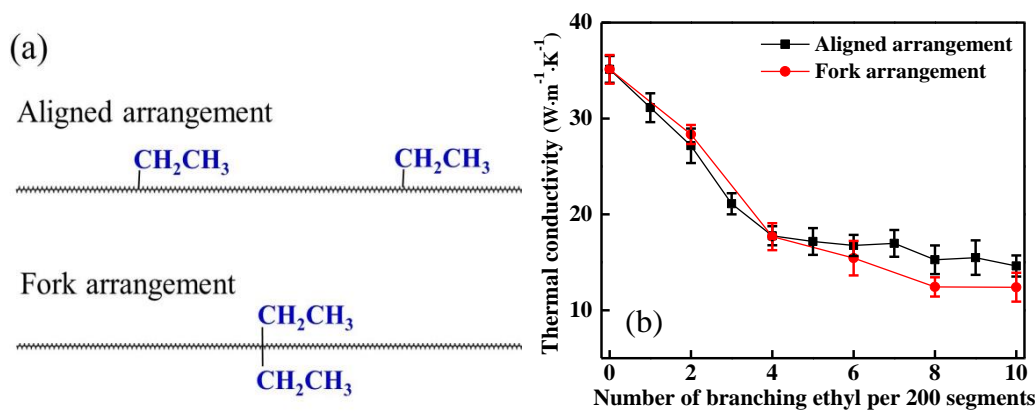

*Fig. 10 TC of a PE chain with different number density of branches: (a) two special branch arrangements (only a part is shown here); (b) TC comparison between two arrangements.*

## 4. Conclusions

It is desirable to further reduce the TC of a polymer for developing a thermal insulation material. In this paper, we take branches into account to probe a way to reduce the TC of a chain. With the RNEMD method applied, the TC of the pristine PE chain and the PE-ethyl chain are simulated and compared. Influences of the chain length, branch arrangements, types and number density of branches are considered. Our results suggest that the branch has a positive effect to decrease the TC of a PE chain. If the number density of ethyl branches becomes larger than 8 ethyl per 200 segments, the TC of a PE-ethyl chain converges to be only about 40% that of a pristine PE chain. This result will not be influenced by the branch arrangements. Different kinds of branching chains lead to a different decrease of thermal conductivities because of the different weights, and a heavy branch leads to a lower thermal conductivity than a light one. This study is expected to provide some fundamental guidance to obtain a polymer with a quite low TC.

**Acknowledgements**

The authors acknowledge the financial support of this work by the National Natural Science Foundation of China (No. 51406224).

**References**

[ 1 ] Li G., Shrotriya V., Yao Y., et al. Manipulating regioregular poly (3-hexylthiophene): [6,6]-phenyl-C61-butyric acid methyl ester blends-route towards high efficiency polymer solar cells [J]. Journal of Materials Chemistry, 2007, 17(30): 3126-3140.

[2] Nie Z, Kumacheva E. Patterning surfaces with functional polymers [J]. Nature Material, 2008, 7(4): 277-290.

[3] Liu C. Recent Developments in Polymer MEMS [J]. Advanced Materials, 2007, 19(22): 3783-3790.

[4] Ryan A. J. Nanotechnology: Squaring up with polymers [J]. Nature, 2008, 456(7220): 334-336.

[5] Bruening M., and Dotzauer D. Polymer films: Just spray it [J]. Nature Material, 2009, 8(6): 449.

[6] Charnley M., Textor M., and Acikgoz C. Designed polymer structures with antifouling-antimicrobial properties [J]. Reactive & Functional Polymers, 2011, 71(3): 329-334.

[7] Han Z., and Fina A. Thermal conductivity of carbon nanotubes and their polymer nanocomposites: A review [J]. Progress in Polymer Science, 2011, 36(7): 914-944.

[8] Singh V., Bougher T. L., Weathers A, et al. High thermal conductivity of chain-oriented amorphous polythiophene [J]. Nature Nanotechnology, 2014, 9(5): 384-390.

[9] Henry A., and Chen G. High Thermal Conductivity of Single Polyethylene Chains Using Molecular Dynamics Simulations [J]. Physical Review Letters, 2008, 101(23): 235502.

[10] Cao B. Y., Li Y. W., Kong J., et al. High thermal conductivity of polyethylene nanowire arrays fabricated by an improved nanoporous template wetting technique[J]. Polymer, 2011, 52(8): 1711-1715.

[11] Henry A., Chen G., Plimpton S.J., et al. 1D-to-3D transition of phonon heat conduction in polyethylene using molecular dynamics simulations [J]. Physical Review B Condensed Matter, 2010, 82(14): 144308.

[12] Jiang J.W., Zhao J., Zhou K., et al. Superior thermal conductivity and extremely high mechanical strength inpolyethylene chains from ab initio calculation [J]. Journal of Applied Physics, 2012, 111(12): 124304.

[13] Shen S., Henry A., Tong J., et al. Polyethylene nanofibres with very high thermal conductivities [J]. Nature Nanotechnology, 2010, 5(4): 251-255.

[14] Liu J., and Yang R. Length-dependent thermal conductivity of single extended

polymer chains [J]. Physical Review B, 2012, 86(10): 104307.

[15] Luo T., Esfarjani K., Shiomi J., et al. Molecular dynamics simulation of thermal energy transport in polydimethylsiloxane (PDMS) [J]. Journal of Applied Physics, 2011, 109(7): 074321.

[16] Robbins A. B., Minnich A. J. Crystalline polymers with exceptionally low thermal conductivity studied using molecular dynamics [J]. Applied Physics Letters, 2015, 107(20): 201908.

[17] Umur A., Gemert M. J. C. V., Ross M. G. Introduction to physical polymer science [M]. Wiley, 1986.

[18] Hu Y., Zeng L., Minnich A. J., et al. Spectral mapping of thermal conductivity through nanoscale ballistic transport.[J]. Nature Nanotechnology, 2015, 10(8): 701

[19] Zeng L., Collins K. C., Hu Y., et al. Measuring Phonon Mean Free Path Distributions by Probing Quasiballistic Phonon Transport in Grating Nanostructures [J]. Scientific Reports, 2015, 5: 17131.

[20] Henry A., and Chen G. Anomalous heat conduction in polyethylene chains: Theory and molecular dynamics simulations [J]. Physical Review B, 2009, 79(14): 144305.

[21] Sasikumar K., and Keblinski P. Effect of chain conformation in the phonon transport across a Si-polyethylene single-molecule covalent junction [J]. Journal of Applied Physics, 2011, 109(11): 114307.

[22] Savin A.V., and Savina O. I. Dependence of the thermal conductivity of a polymer chain on its tension [J]. Physics of the Solid State, 2014, 56(8): 1664-1672.

[ 23 ] Liao Q., Zeng L., Liu Z., et al. Tailoring Thermal Conductivity of Single-stranded Carbon-chain Polymers through Atomic Mass Modification [J]. Scientific Reports, 2016, 6: 34999.

[24] Ma H., and Tian Z. Effects of polymer topology and morphology on thermal transport: A molecular dynamics study of bottlebrush polymers [J]. Applied Physics Letters, 2017, 110(9): 091903.

[25] Zhang T., Wu X., and Luo T. Polymer Nanofibers with Outstanding Thermal Conductivity and Thermal Stability: Fundamental Linkage between Molecular Characteristics and Macroscopic Thermal Properties [J]. Journal of Physical Chemistry C, 2014, 118(36): 21148-21159.

[26] Zhang L., Ruesch M., Zhang X., et al. Tuning thermal conductivity of crystalline polymer nanofibers by interchain hydrogen bonding [J]. RSC Advances, 2015, 5(107):87981-87986.

[ 27 ] Zhang T., and Luo T. Morphology-influenced thermal conductivity of polyethylene single chains and crystalline fibers [J]. Journal of Applied Physics, 2012, 112(9): 1571.

[ 28 ] Ma H., and Tian Z. Effects of polymer chain confinement on thermal conductivity of ultrathin amorphous polystyrene films [J]. Applied Physics Letters, 2015, 107(7): 073111.

[29] Sun H. COMPASS: An ab Initio Force-Field Optimized for Condensed-Phase Applications overview with Details on Alkane and Benzene Compounds [J]. Journal of Physical Chemistry B, 1998, 102(38): 7338-7364.

[30] Sun H., Ren P., and Fried J. R. The COMPASS force field: parameterization and

validation for phosphazenes [J]. Computational & Theoretical Polymer Science, 1998, 8(1-2):229-246.

[31] Rigby D., Sun H., and Eichinger B. E. Computer simulations of poly (ethylene oxide): force field, pvt diagram and cyclization behavior [J]. Polymer International, 1997, 44(3): 311-330.

[32] Ma B., Riggs J. E., and Sun Y. P. Photophysical and Nonlinear Absorptive Optical Limiting Properties of [60]Fullerene Dimer and Poly[60]fullerene Polymer [J]. The Journal of Physical Chemistry B, 1998, 102(31), 5999-6009.

[33] Li C., Choi P., and Sundararajan P. R. Simulation of chain folding in polyethylene: A comparison of united atom and explicit hydrogen atom models [J]. Polymer, 2010, 51(13): 2803-2808.

[34] Luo T., and Lloyd J. R. Enhancement of Thermal Energy Transport Across Graphene/Graphite and Polymer Interfaces: A Molecular Dynamics Study [J]. Advanced Functional Materials, 2012, 19(12): 587-596.

[35] Nose S. A unified formulation of the constant temperature molecular dynamics methods[J]. Review of Faith & International Affairs, 1984, 81(2):25-31.

[36] Hoover W. G. Canonical dynamics: Equilibrium phase-space distributions.[J]. Physical Review A, 1985, 31(3):1695-1697.

[37] Müllerplathe F. A simple nonequilibrium molecular dynamics method for calculating the thermal conductivity [J]. Journal of Chemical Physics, 1997, 106(14): 2878-2891.

[38] Huang C., Wang Q. and Rao Z. Thermal conductivity prediction of copper hollow nanowire [J]. International Journal of Thermal Sciences, 2015, 94: 90-95.

[39] Lee B., and Richards F. M. The interpretation of protein structures: estimation of static accessibility. Journal of Molecular Biology, 1971, 55 (3): 379-400.

[40] Connolly M. L. Solvent-accessible surfaces of proteins and nucleic acids. Science, 1983, 221: 709.

[41] Turney J. E., McGaughey A. J. H., and Amon C. H. Assessing the applicability of quantum corrections to classical thermal conductivity predictions [J]. Physical Review B, 2009, 79(22): 224305.

[42] Moreland J. F. The disparate thermal conductivity of carbon nanotubes and diamond nanowires studied by atomistic simulation [J]. Microscale Thermophysical Engineering, 2004, 8(1): 61-69.

[43]Maiti A., Mahan G. D., and Pantelides S. T. Dynamical simulations of nonequilibrium processes-Heat flow and the Kapitza resistance across grain boundaries [J]. Solid state communications, 1997, 102(7): 517-521.

[44] Lukes J. R., and Zhong H. Thermal conductivity of individual single-wall carbon nanotubes. Journal of Heat Transfer, 2007, 129(6): 705-716.

[45] Lee Y. H., Biswas R., Soukoulis C. M., Wang C. Z., Chan C. T., and Ho K. M. Molecular-Dynamics Simulation of Thermal Conductivity in Amorphous Silicon [J]. Physical Review B, 1991, 43(8): 6573–6580.

[46] Ni B., Watanabe T., and Phillpot S. R. Thermal transport in polyethylene and at polyethylene-diamond interfaces investigated using molecular dynamics simulation [J]. Journal of Physics: Condensed Matter, 2009, 21(8): 2820-2823.

[47] Hu G. J., Cao B. Y. and Li Y. W. Thermal Conduction in a Single Polyethylene

Chain Using Molecular Dynamics Simulations[J]. Chinese Physics Letters, 2014, 31(8): 119-122.

[48] Termentzidis K., Merabia S., Chantrenne P., and Keblinski P. Cross-plane thermal conductivity of superlattices with rough interfaces using equilibrium and non-equilibrium molecular dynamics. International Journal of Heat and Mass Transfer, 2011, 54(9), 2014-2020.

[49] Landry E. S., Hussein M. I., and McGaughey A. J. H. Complex superlattice unit cell designs for reduced thermal conductivity. Physical Review B, 2008,77(18), 184302.